\documentclass[preprint,a4paper]{aastex}

\usepackage{graphics,epsf}
\usepackage{amsmath}                
\usepackage{amsfonts}               
\usepackage{amssymb}                
\usepackage{epsfig}
\usepackage{rotating}

\def \cm{~\rm{cm}}
\def \s{~\rm{s}}
\def \km{~\rm{km}}

\def \K{~\rm{K}}
\def \g{~\rm{g}}

\def \AU{~\rm{AU}}

\def \yr{~\rm{yr}}

\def \days{~\rm{day}}

\begin{document}

\title{ACCRETION OF DENSE CLUMPS IN THE PERIASTRON PASSAGE OF ETA CARINAE}

\author{Muhammad Akashi \altaffilmark{1}$^*$, Amit Kashi\altaffilmark{2}, and Noam Soker\altaffilmark{1}}
\altaffiltext{1}{Department of Physics, Technion$-$Israel Institute of Technology,
Haifa 32000, Israel; akashi@physics.technion.ac.il;  soker@physics.technion.ac.il.}
\altaffiltext{2}{Department of Physics and Astronomy, University of Nevada, Las Vegas,
4505 S. Maryland Pkwy, Las Vegas, NV, 89154-4002, USA; kashia@physics.unlv.edu.}
\begin{abstract}

We perform 3D hydrodynamical numerical simulations of the winds interaction process in the massive binary system
$\eta$ Carinae, and find the secondary star to accrete mass from the dense primary wind close to periastron passage.
This accretion is thought to result in the spectroscopic event and X-ray minimum observed in the system every revolution.
In this study we limit ourselves to explore the role of clumps in the primary wind in
triggering the accretion process.
We include the gravity of the secondary star and the orbital motion starting 19 days (90 degrees) before periastron passage.
The accretion process is triggered by dense clumps that cannot be decelerated by the ram pressure of the secondary wind.
The dense clumps are formed by instabilities in the thin dense shell formed by the shocked primary wind gas.
We explore the role of the numerical viscosity and some physical parameters on the initiation of the accretion process,
and explain the unique properties of $\eta$ Car that allow for the periastron accretion process to occur.
The accretion starts about a week before periastron passage, as is required to explain the several weeks long X-ray minimum.

\end{abstract}


\section{INTRODUCTION}
\label{sec:intro}

The very massive binary system $\eta$ Car is one of the most studied binary systems, and for good reasons.
With a rare type of primary --- a luminous blue variable (LBV) star,
and a secondary in an unusual highly eccentric orbit ($e\simeq0.9$--$0.95$; e.g., Hillier et al. 2006)
it is prone to effects that are rarely seen in other systems.
The most known of which is the 1838--1858 Great Eruption which created the Homunculus nebula,
whose mass is estimated to be $10$--$40 {\rm M_\odot}$
(Gomez et al. 2006, 2009; Smith et al. 2003; Smith \& Ferland 2007).

The two stellar winds collide and create a conical shell structure.
The source of the hard X-ray emission is the shocked
wind of the secondary star (Corcoran et al. 2001; Pittard \& Corcoran 2002; Akashi et al. 2006).
Some major puzzles are connected to the process that causes the 4--10 weeks long decrease in the X-ray
emission during each periastron passage (Corcoran 2005; Hamaguchi et al. 2007),
accompanied by a decrease in the equivalent width and/or significant changes in some properties of many
emission and absorption lines in what is termed the `spectroscopic event' (e.g., Davidson et al. 2005; Smith 2005;
Martin et al. 2006a,b; Nielsen et al 2007; Damineli et al. 2008; Mehner et al. 2010, 2011).
Kris Davidson was probably the first person to explain that the spectroscopic events are not simply eclipses (Davidson 1999),
and to bring up the idea of shock instabilities and shock-breakup in the colliding winds (Davidson 2002).
The long duration of the deep X-ray minimum imposes strong constrains on the properties of the binary system
(Kashi \& Soker 2009b,c)

In previous papers it was suggested that close to periastron passage of $\eta$ Car accretion onto the companion occurs.
The pioneering paper describing this possibility was Soker (2005).
The idea of accretion onto the companion was introduced to explain the cyclic spectroscopic event and X-ray minimum.
Soker (2005) suggested that cold and dense clumps of size $\ga 0.001 D_2$ will be accreted by the
secondary, where $D_2$ is the distance between the conical shell and the secondary.
Such clumps, Soker (2005) suggested, might form a few weeks before periastron passage.
It was also suggested that only few of those clumps are enough to shut down the secondary wind.
As there is no secondary wind, there is no collision with the primary wind, no shocks, and no X-ray emission.
One of the uncertainties in this study involves the exact velocity profile and terminal velocities of the two winds.
In particular, the radiation from one star can influence
the velocity profile of the wind from the other star before the winds collide (Gayley et al. 1997).

The accretion model was discussed in a series of papers (e.g., Soker 2005b; Akashi et al. 2006; Soker \& Behar 2006;
Kashi \& Soker 2009a).
Kashi \& Soker (2009b) constructed a quantitative model to illustrate the
accretion process, under the assumption that for several weeks near periastron passages
the secondary star accretes mass from the slow dense wind blown by the primary star.
As an approximation to the complicated accretion process, two accretion processes were taken into
consideration in an analytic calculation, Bondi-Hoyle-Lyttleton accretion (wind accretion), and Roche lobe overflow.
Kashi \& Soker (2009a) found that the secondary accretes few $\times 10^{-6} {\rm M_\odot}$ close to periastron.
It was further suggested that a thick accretion belt is formed around the secondary, and its
presence accounts for several observational properties across the spectroscopic event.
In Kashi \& Soker (2009a) we also stressed the point that accretion cannot be prevented by the radiation pressure of the secondary.

The accretion model can account for the variations in the durations of the X-ray
minimum observed in different cycles.
Kashi \& Soker (2009b) attribute the variation in the length of the X-ray minimum to variations in the velocity or mass loss
rate of the primary wind, that cause variation in the recovery time of the secondary wind after the minimum.
Having a lower mass loss rate is also supported by observations of weakening of emission lines originationg in the primary LBV, especially the H$\alpha$ and the He~I $\lambda6680$ lines (Mehner at al. 2010). 
In the same manner that lower mass loss from the primary star may have hasted the recovery of the secondary wind after the 2009 spectroscopic event, 
a higher mass loss rate would delay the recovery of the secondary wind.

Early numerical simulations of the two winds collision in $\eta$ Car did
not include the gravity of the secondary star,
and could not obtain an accretion (Pittard et al. 1998;
Pittard \& Corcoran 2002; Okazaki et al. 2008; Parkin et al. 2009).
Numerical simulations that did include the gravity of the stars
(Pittard 1998; Stevens \& Pollock 1994) had parameters that are not
relevant to $\eta$ Car, and did not consider accretion.

Akashi \& Soker (2010) were the first to performed 3D numerical simulations that include gravity with parameters
appropriate to the $\eta$ Car binary system.
However, they did not include the orbital motion.
Their main finding was that accretion is initiated by dense clumps that are formed in the winds collision region.
Parkin et al. (2011) presented high quality 3D adaptive mesh refinement hydrodynamical simulations of the winds in $\eta$ Car,
but found no accretion when orbital motion was included.
In that paper and a previous one (Parkin et al. 2009) they discuss a ``collapse'' of the shocked wind onto the secondary
at periastron passage.
There is no discussion of what happens to the primary wind during that time, namely whether
it is accreted onto the secondary or immediately expelled from the secondary when reaching it.
If it is not expelled then the conclusion should be that accretion onto the secondary occurs,
as in the model invoked by Soker (2005).

The models of Parkin et al. (2011) do not reach the same minimum in X-ray flux as observed (Corcoran 2005, 2010).
Parkin et al. (2011) considered it a puzzling result as their previous models were able to reproduce
the low flux level at periastron via an eclipse of the X-ray emission region (Okazaki et al. 2008; Parkin et al. 2009).
The explanation of Parkin et al. (2011) to this anomaly is that their model had two unwanted components
that contributed to the X-ray flux, emission from the trailing arm of the wind collision region and
a weaker contribution from the downstream gas in the leading arm.
We believe that the higher flux obtained in the models of Parkin et al. (2011) is not surprising,
as models with no accretion fail to reproduce the X-ray minimum (see discussion in Kashi \& Soker 2009c).
As a side comment, we note that the calculation of the X-ray emission far from periastron might suffer from
large uncertainties because of the presence of multi-scale stochastic wind clumps in the primary wind (Davidson 2002; Moffat \& Corcoran 2009). 

It is important to emphasize the major role of clumps in the accretion process (Soker 2005).
In Kashi \& Soker (2009a), models with different acceleration profiles of the primary wind
(different values of the beta parameter) were examined.
Higher values of beta (larger wind acceleration zone) which are associated with clumpy wind,
lead to accretion of a larger amount of material onto the secondary.

In this paper we follow the winds interaction near periastron passage, and include the orbital motion.
We find that instabilities lead to the formation of dense clumps,
that in turn lead to accretion onto the companion near periastron passage.
The study is focused on the formation and role of clumps in the primary wind and their role in starting the accretion process. The study of the formation and acceleration of the two winds, as well as of preexisting clumps in the primary wind will be the subject of a forthcoming paper.
In a still later paper the exact interaction of the accreted  clumps with the secondary wind and radiation will be studied.
In section \ref{sec:numeric} we present the numerical code and parameters.
Section \ref{sec:results} presents the results of our different runs and the physical processes involved.
We summarize in section \ref{sec:summary}.

\section{NUMERICAL SET UP}
\label{sec:numeric}

There are some uncertainties as to the binary
parameters of $\eta$ Car (Kashi \& Soker 2009a).
We here take the commonly accepted parameters, as listed in papers listed in section 1.
The stellar masses are $M_1 = 120 {\rm M_\odot}$ and $M_2 = 30 {\rm M_\odot}$,
the eccentricity is $e = 0.9$, and the orbital
period $P = 2024~$ days, hence the semi-major axis
is $a = 16.64 \AU$ and periastron occurs at $r_p = 1.66 \AU$.
The mass loss rates and velocities of the winds are
$\dot M_1= 3 \times 10^{-4} {\rm M_\odot} \yr^{-1}$ and
$\dot M_2(t)= 10^{-5} {\rm M_\odot} \yr^{-1}$, and $v_1 = 500 \km \s^{-1}$
and $v_2 = 3000 \km \s^{-1}$, respectively.

The simulations are performed with Virginia
Hydrodynamics-I (VH-1), a high
resolution multidimensional astrophysical
hydrodynamics code developed by
John Blondin and co-workers (Blondin et al. 1990;
Stevens et al., 1992; Blondin 1994).
We have added radiative cooling to the code at
all temperatures of $T > 10^4 \K$.
Namely, the radiative cooling is set to zero
for temperatures of $T < 10^4 \K$,
but we allow adiabatic cooling.
Radiative cooling is carefully treated near contact
discontinuities, to prevent large temperature
gradients from causing unphysical results.
The cooling function $\Lambda (T)$ (for solar abundances) is taken
from Sutherland \& Dopita (1993; their table 6).
Gravity by the secondary star is included, as this
is the sole issue of this study.
For the primary we neglect both gravity and radiation pressure that accelerate the wind.
We start by placing the secondary at $90^\circ$ from periastron,
corresponding to an orbital separation of $3.17 \AU$, and 19.1 days
before periastron ($t=-19.1$~day).
We impose the two winds in the grid, and let
the flow reach a steady state with no gravity yet.
Only then we turn on gravity, and let the flow reach a
new steady state with gravity included.
Only after the new steady state is reached we start the orbital motion of the secondary star
(except in run A5 in which the secondary does not move).
In our grid the position of he primary star is fixed, and the secondary move in its orbital trajectory.
Making the orbit around the primary star rather than around the center of mass is adequate enough for our goals,
the main one to understand
the basic role of the secondary gravity and of dense clumps.
As we are interested only in the accretion process, we limit the calculation to the
19.1 days prior to periastron passage, when the binary vector radius is at
$90^\circ$ to periastron angle, and until periastron passage.

We perform the numerical simulations in the Cartesian geometry $(x,y,z)$ mode
of the code (a 3D calculation).
In most runs there are 119 grid points along each axis.
The size of all cells is the same,
but the length along the three axes is not equal, i.e., the cells are not cubes, but rather boxes.
The computational volume includes only one side of the orbital plane.
To confirm the adequacy of the resolution, we run a case with 189 cells along each axis.
We found the differences between the low and high resolution runs to be small, and therefore
consider the lower resolution runs to be adequate for representing the colliding winds.

In most simulations, the grid size is $2.12 \AU \times 7.45 \AU \times 7.45 \AU$,
along the $x,y,z$ directions, respectively, where the secondary is along the positive $x$-axis at periastron.
At the location of each star we inject its appropriate wind.
The primary wind was injected from a cube of size $15\times15\times15$ cells around it.
The secondary wind was injected from a box of size $11\times7\times7$ cells around it.
This prevents the primary wind to reach a distance closer than 6 cells (along the $x$ axis) and 4 cells
(along the y and z axis) from the center of the secondary .
The boundary conditions of the simulation box are outflows at the $6$ sides of the box.

\section{RESULTS}
\label{sec:results}

As our main goal is to explore the role of some physical parameters on the accretion process near periastron passages,
we will divide the results according to the role of these parameters. We will concentrate on the flow structure
as periastron passage is approached, as this is the stage when accretion starts.
We cannot resolve the acceleration zone of the secondary wind, and hence cannot deal with the flow of accreted gas
to the secondary surface.
Therefore, the condition for accretion is taken to be the presence of a dense
clump near the numerical injection region (a box) of the secondary wind (Akashi \& Soker 2010).
By `near' we take a distance of one cell from the injection box to the outskirts of the dense clump.
As will be evident in this section, the presence of accretion by this criterion is clear from the density contour maps.
Soker (2005) have shown that the secondary wind and radiation pressure cannot prevent the accretion
of dense clumps as obtained here when they come close to the secondary.

We note that some papers, e.g., Teodoro et al. (2011; also Damineli et al. 2008) mention a process of
``collapse of the primary wind onto the secondary star''.
We are not sure what `collapse' means.
As the primary wind material starts its journey not being gravitational bound to the secondary star
(but rather to the primary star), and later when the clumps
come close to the secondary they are gravitational bound to the secondary star, we think the appropriate
term is `accretion'.

The results will be presented as 2D cuts through the grid, showing the density,
temperature, and velocity maps for the different models and at different times.
The time will be given as time before periastron passage.
The different runs are summarized in Table 1.
\begin{table}

Table 1: Parameters of the runs.

\bigskip
\begin{tabular}{|l|c|c|c|c|c|l|}
\hline
 Run & $M_1 ({\rm M_\odot})$ & $ M_2 ({\rm M_\odot})$ & Grid points & Fig. & Accretion & Comments \\
\hline
A1 &    0 &  0 & $119^{3}$ & 1     & No  & No Gravity \\
\hline
A2 &  120 & 30 & $119^{3}$ & 2     & Yes & The standard run \\
\hline
A3 &  120 & 30 & $189^{3}$ & 3,4,5 & Yes & High resolution \\
\hline
A4 &  120 & 30 & $119^{3}$ & 6,7   & No  & High artificial viscosity \\
\hline
A5 &  120 & 30 & $119^{3}$ & 8     & No  & Static secondary star\\
\hline
A6 &  120 & 30 & $189^{3}$ & 9     & Yes & Larger grid size, high resolution \\
\hline

\end{tabular}

\footnotesize
\bigskip
For all runs the stellar mass loss rates and velocities are: 
$\dot M_1= 3 \times 10^{-4} {\rm M_\odot \yr^{-1}}$ and
$\dot M_2= 10^{-5} {\rm M_\odot} \yr^{-1}$, $v_1 = 500 \km \s^{-1}$
and $v_2 = 3000 \km \s^{-1}$.
\normalsize
\end{table}

\subsection{No accretion without gravity}
\label{sec:gravity}
\begin{figure}
 \centering

   \includegraphics[scale=0.85]{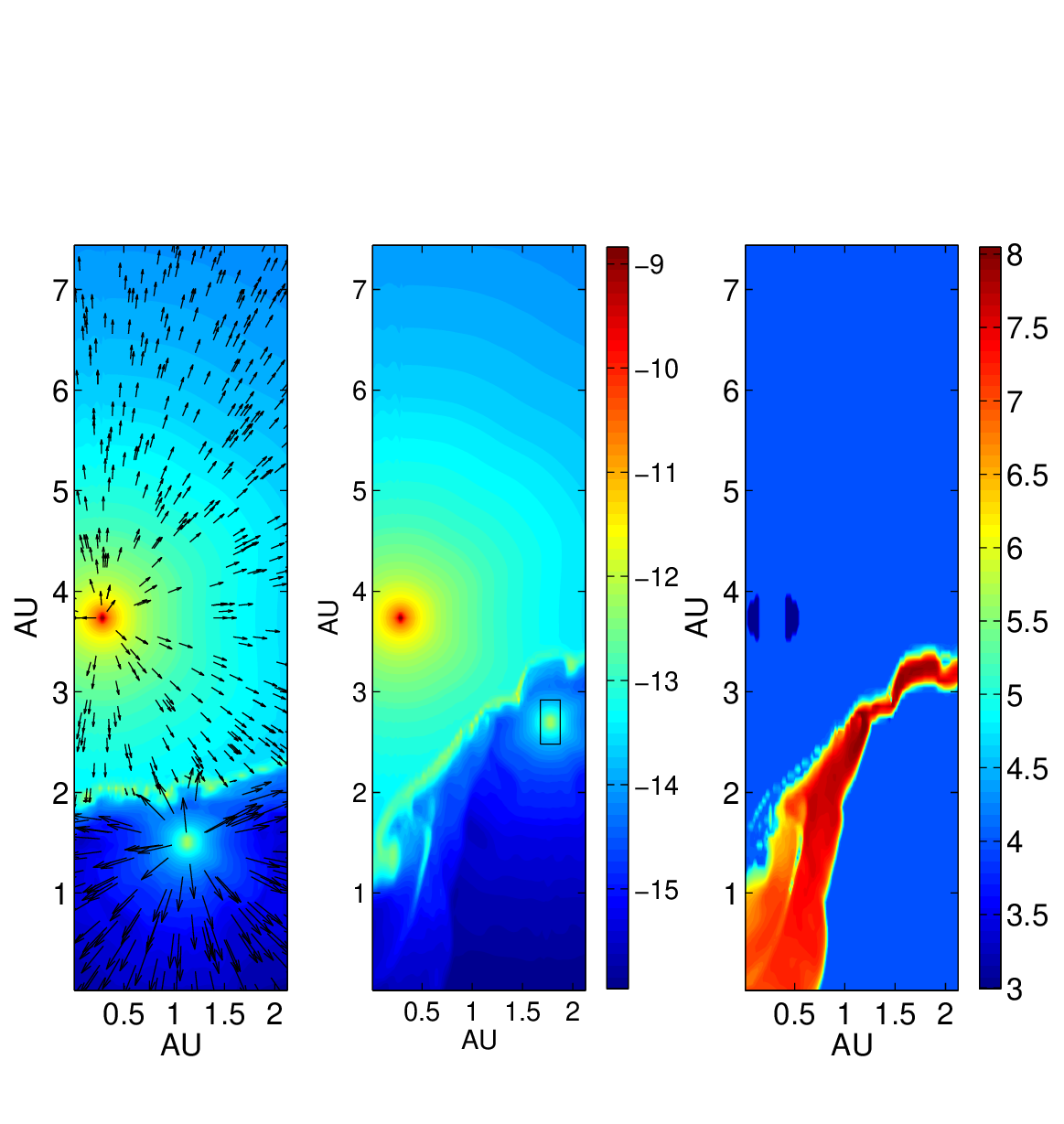}

        \caption{Results for run A1 in the $xy$ plane (the equatorial plane).
        Left: The Density map with velocity arrows at the time $t=-12.2~{\rm day}$, where $t=0$ is periastron passage.
        The velocity arrows represent the flow velocity at the location of the arrow's foot.
        It is presented with four arrow sizes
        ($0 \km \s^{-1}<v<750 \km \s^{-1}$, $750 \km \s^{-1}<v<1500 \km \s^{-1}$, $1500 \km \s^{-1}<v<2250 \km \s^{-1}$, and $2250 \km \s^{-1}<v<3000 \km \s^{-1}$).
        Middle: the density map at $t=-5.2~{\rm day}$.
        Right: temperature map at $t=-5.2~{\rm day}$.
        The bar on the right gives the log scale of density and temperature, and color-coded in units of
        $ \g \cm^{-3} $ and  $\K$, respectively.}
   \label{fig:A1}
     \end{figure}

We start by showing that without gravity no accretion occurs
(see also Akashi \& Soker 2010), although dense clumps do form.
In Fig. \ref{fig:A1} we present the density (first two panels) velocity (first panel) and
temperature (third panel) in the orbital plane for run A1 that has no gravity.


The classical structure of the winds interaction region is seen here (Stevens et al. 1997; Pittard et al. 1998;
Pittard \& Corcoran 2002; Pittard \& Corcoran 2002; Okazaki et al. 2008; Parkin et al. 2009; Parkin et al. 2011).
The shock of the secondary wind is located where there is a huge temperature jump. A hot gas region (hot bubble)
is formed in the post-secondary wind region.
The primary wind shock is located at one boundary of the thin dense shell. The other boundary of
the dense thin shell touches the hot bubble. This thin shell suffers from the thin shell instability (Vishniac 1983, 1984),
and dense clumps are formed and the thin shell wiggles. 

We follows the orbit from $t=-19.1 \days$ to periastron passage, and at no time we find a clump that cames close to the
secondary wind injection region.
This is taken as no accretion occurs in this case, although the clumps come closer to the secondary than
the expected distance of an unperturbed winds interaction region (smooth thin shell).
The third panel is a temperature map, in which we see the shock heating the
gas up to a temperature of $\sim 10^{8} \K$.
This teaches us that if accretion does not occur the
hard X-ray emission continues beyond periastron passage, contrary to observation (Corcoran 2005; Corcoran et al. 2010).

\subsection{Accretion}
\label{sec:accretion}

\begin{figure}
 \centering

   \includegraphics[scale=0.95]{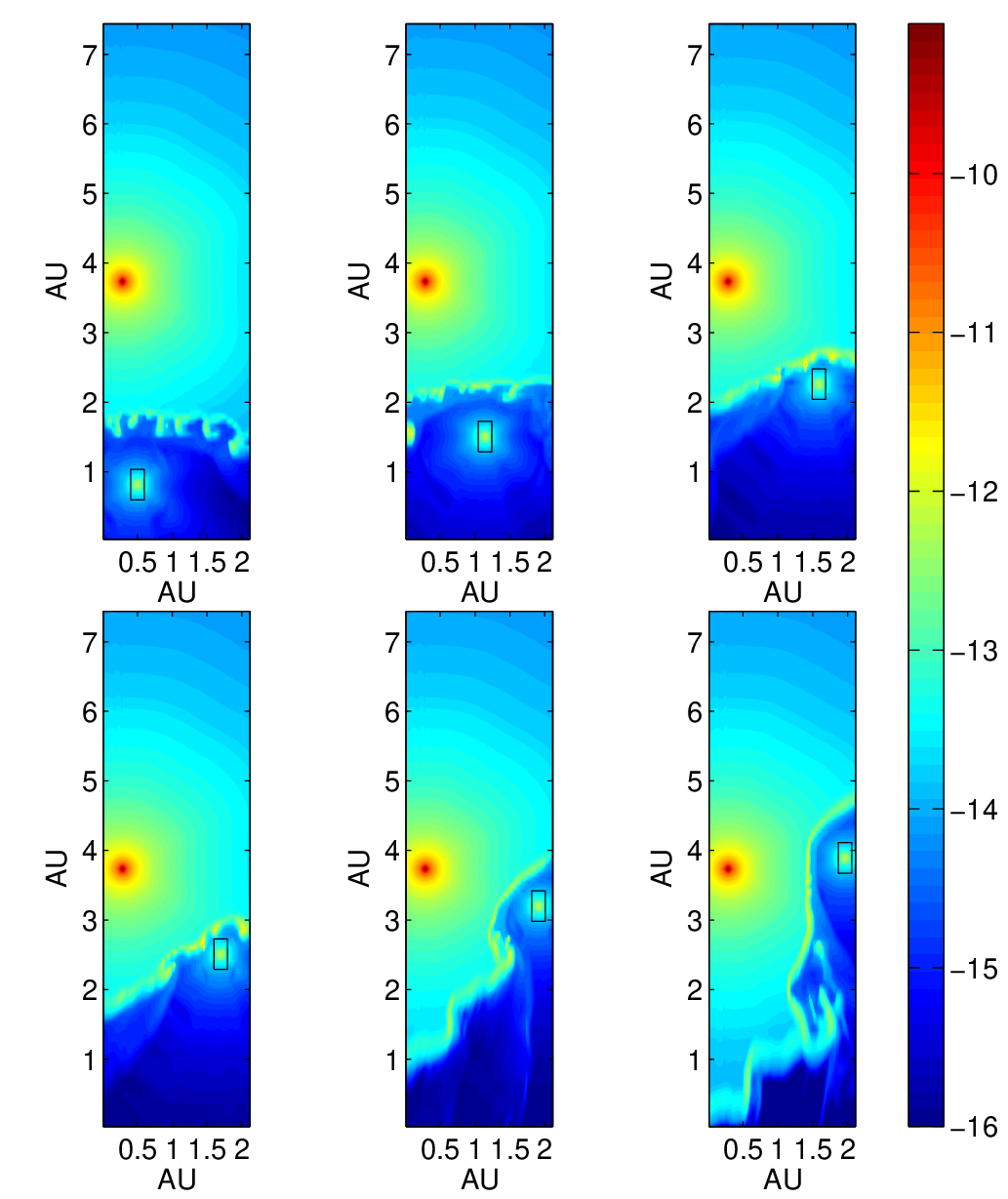}

        \caption{The density of run A2 in the $xy$ plane at six different times: Clockwise from upper left in days relative to periastron passage:
         (1) -17.9 (2) -12.2 (3) -7.5  (4) -6.4 (5) -2.9 (6) 0.6;
        The black rectangle is the cut of the $xy$ plane through the secondary wind injection box.
        A dense clump touching the injection region is seen in the 4th panel; this is taken as indication for accretion.
        In reality, the flow depicted in the last two panels does not occur because accretion starts before that and
        shuts off the secondary wind. We show the flow at periastron passage to emphasize that without accretion
        the X-ray emission at periastron passage is much larger than observed.
        The bar on the right gives the log scale of density, and color-coded in
        units of $ \g \cm^{-3} $.}
   \label{fig:Standard1}
     \end{figure}

Our standard run A2 has the same parameters as run A1, but gravity is added to the secondary star.
In Fig. \ref{fig:Standard1} we present the density in the equatorial plane at six different times.
The secondary wind injection region is a box, and its cut on the $xy$ plane is a rectangle.
This rectangle is marked in all panels of the figure.
In the first panel of Fig. \ref{fig:Standard1} we can clearly see the formation of instabilities
within the thin dense shell.
As the stars approach periastron the dense clumps are seen closer
to the secondary star.
In panel 4 of Fig. \ref{fig:Standard1} a clump touching the rectangle is clearly seen: accretion occurs in this case,
and it is triggered by the clump.
After accretion starts it will substantially disturbed the secondary wind.
Our code cannot handle the flow anymore.
We still present two panels at later times to show that if accretion is not considered, a hot bubble is formed
and strong X-ray emission is expected, contrary to observations. Namely, the flow depicted in the last two panels
does not occur in $\eta$ Car, and shown here for pedagogical reasons (same for some later figures).

\begin{figure}
 \centering

   \includegraphics[scale=0.95]{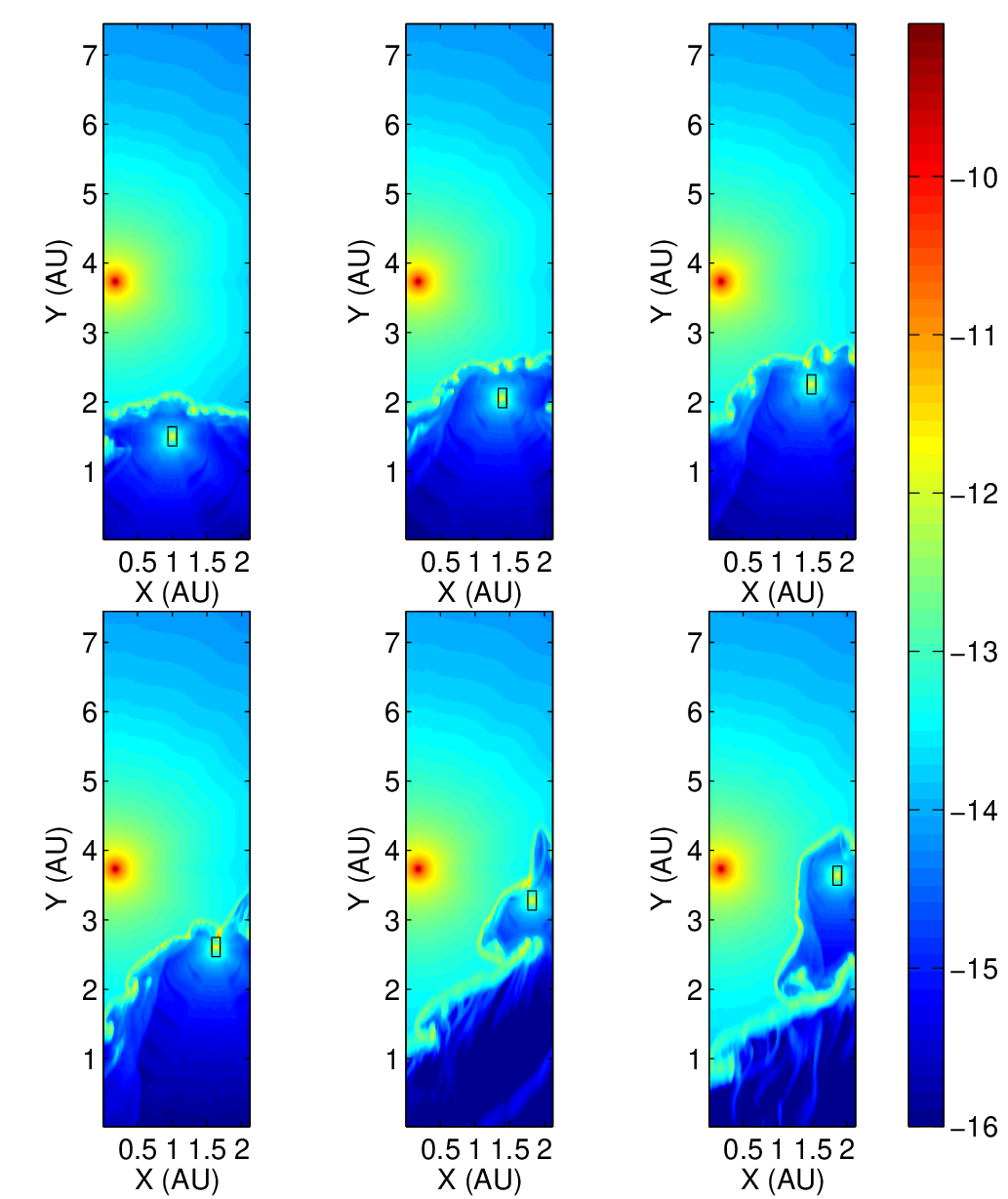}

        \caption{A high resolution run, with $189^{3}$ cells, for the same conditions as in Fig. \ref{fig:Standard1}, and
at the times: (day) (1) -12.3 (2) -8.6 (3) -7.5 (4) -5.8 (5) -2.3 (6) -0.6.
Here as well, the flow depicted in the last two panels does not occur in $\eta$ Car because accretion starts before then,
as evident from the clump seen accreted in panel 4.}

   \label{fig:Highres1}
     \end{figure}

\begin{figure}
 \centering

   \includegraphics[scale=0.95]{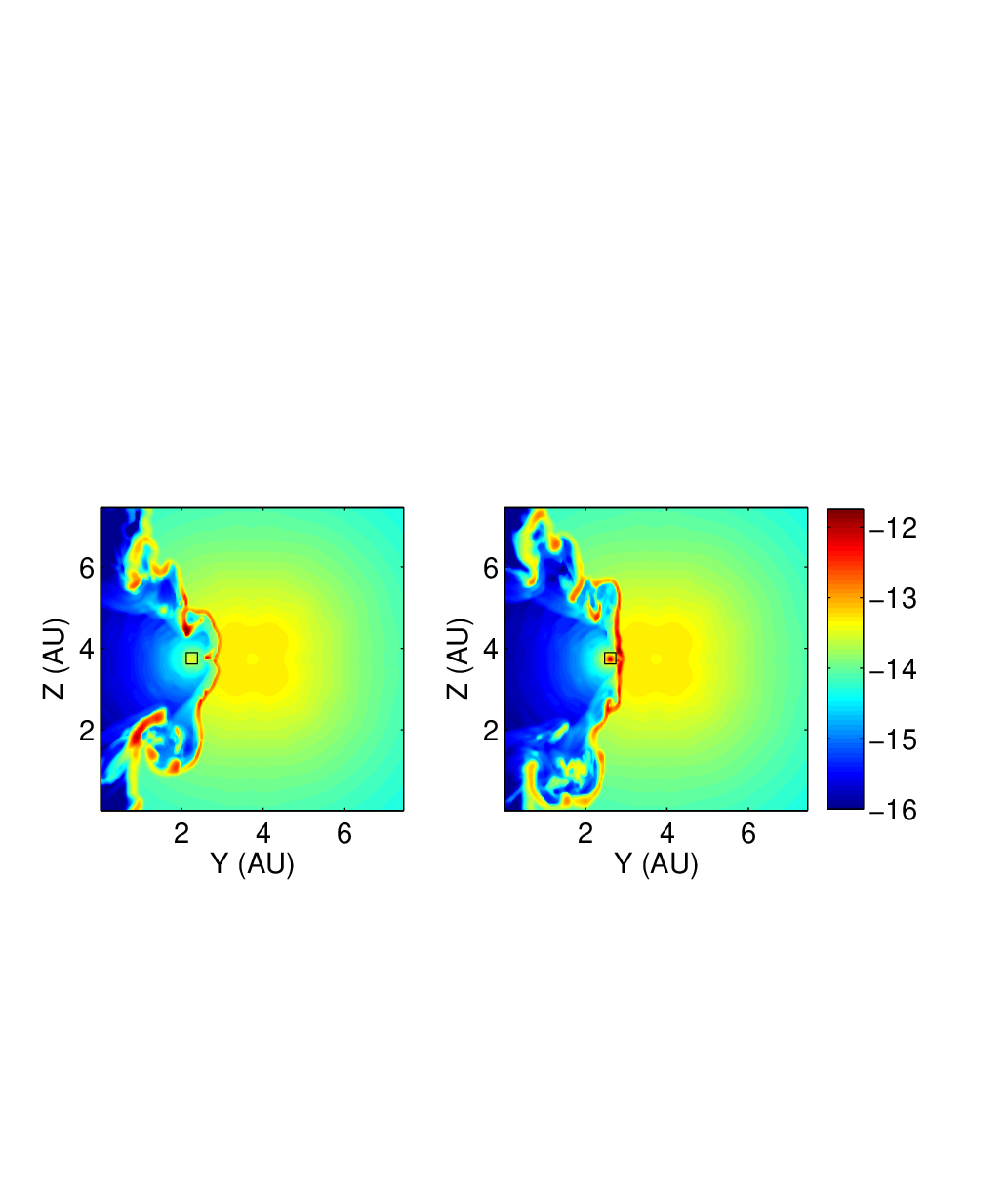}

        \caption{Similar to Fig. \ref{fig:Highres1}, but the density map is shown in the $yz$ plane through the secondary
        star. The two panels correspond to panels 3 ($t=-7.5$) and 4 ($t=-5.8$)in Fig. \ref{fig:Highres1}.
        }
   \label{fig:Highres2}
     \end{figure}

To reveal how close the clump can get, we conduct run A3 having the same parameters as run A2 but
at a high resolution. The secondary wind injection box has the same number of cells, hence it is smaller.
As can be see in the fourth panel of Fig. \ref{fig:Highres1} a dense clump touches the wind's injection box,
and comes very close to the secondary star. This is a clear indication for accretion.
To further demonstrate this, in Fig. \ref{fig:Highres2} we present the density map in a plane
parallel to the $yz$ plane and through the secondary star, at two times corresponding to
panels 3 and 4 in Fig. \ref{fig:Highres1}. In the second panel not only the dense clump in the equatorial plane
is touching the injection zone of the secondary wind, but along the entire hight of the box (the z axis) dense
primary wind material is touching the box. This is a clear indication for a major accretion process.
Soker (2005; eq. 14 there) gives the condition on the size of a blob for the secondary stellar gravity to overcome the
ram pressure of the secondary wind. This is the accretion condition.
The clump seen in the fourth panel of Fig. \ref{fig:Highres1} is elongated, with a half length of $R_b \simeq 0.03 \AU$,
and a density of $\sim  10^{-11}$, located at a distance of $r_2 \simeq 0.25 \AU$ from the secondary.
Such a blob just obeys the accretion condition against the ram pressure of the secondary wind (eq. 14 in Soker 2005).
Due to the lower resolution calculation presented in Fig. \ref{fig:Standard1}, the dense blobs there don't obey the accretion condition.
We believe that a higher resolution calculation that gives more accurate results will make the blob denser,
hence obeying the accretion condition with a larger margin. 

\begin{figure}
 \centering

   \includegraphics[scale=1]{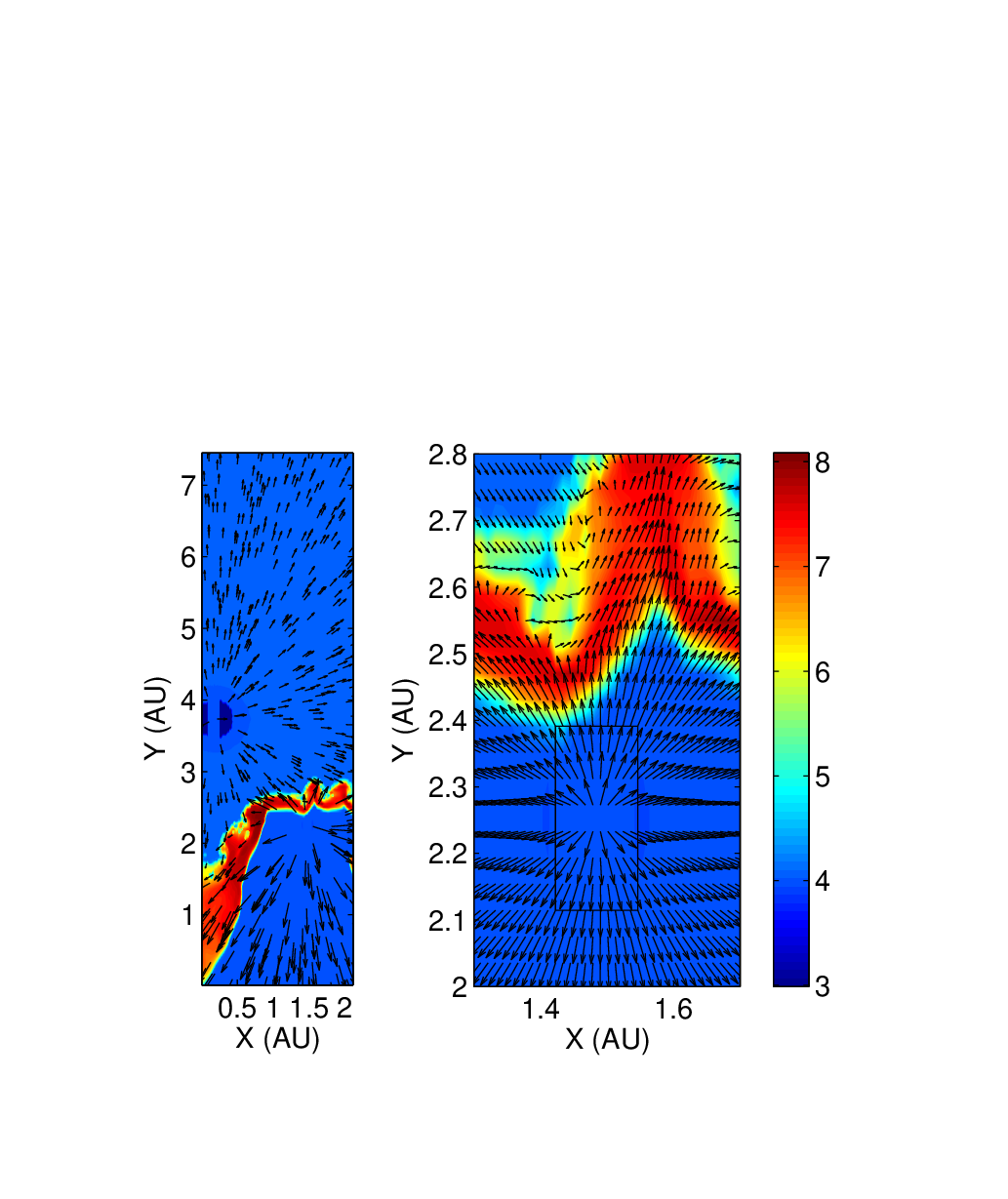}

        \caption{The velocity and temperature map in the $xy$ plane at $t=-7.5 ~{\rm day}$
         (corresponding to panel 3 in Fig. \ref{fig:Highres1}).
         The second panel focuses on the region near the clump.
         Velocity is presented with four arrow sizes
         ($0 \km \s^{-1}<v<750 \km \s^{-1}$, $750 \km \s^{-1}<v<1500 \km \s^{-1}$, $1500 \km \s^{-1}<v<2250 \km \s^{-1}$,
         and $2250 \km \s^{-1}<v<3000 \km \s^{-1}$). Note that the arrow sizes are different than in Fig \ref{fig:A1}.
         The velocity in the right panel is calculated relative to the secondary.
         It is evident that the clump has a velocity component towards the secondary; accretion is about to begin.
         }
   \label{fig:Highres3}
     \end{figure}

In Fig. \ref{fig:Highres3} we show the temperature and velocity maps at $t=-7.5 ~{\rm day}$
(corresponding to panel 3 in Fig. \ref{fig:Highres1}).
In the second panel we focus on the clump, and the velocity
is presented relative to the secondary star rather than to the numerical grid.
It is evident that the clump has a velocity component towards the secondary and that it is about to be accreted onto the secondary.
The size of our clump in Fig 5. is $\sim 0.03 \AU$ while the distance between the conical shell and the secondary
at this phase ($t=-7.5 ~{\rm day}$) is $\sim 0.3 \AU$.
This is consistent with Soker (2005) estimate that cold and dense clumps of size $\ga 0.001 D_2$ will be accreted by the secondary.

In these runs we observe an interesting phenomena, that was noticed and discussed already by Parkin et al. (2011).
Very close to periastron passage there are no clumps (as stated above, the flow there does
not occur in $\eta$ Car, as accretion stars earlier).
If accretion does not start more than few days prior to periastron passage and shuts-off the secondary wind,
accretion will not start at periastron either.
As periastron is approached the stagnation point of the colliding winds moves to the side of the
direction of motion of the secondary, and the instabilities have no time to
develop and form dense clumps near the secondary star.
This is not a problem in the case of $\eta$ Car, because as we show here, accretion starts at $t=- 8~{\rm day}$
when plenty of clumps are formed and can start the accretion process.
We expect that the accretion will shut-off the secondary wind, and will continue as long as the accretion
rate is sufficiently high (Kashi \& Soker 2009a).
In a forthcoming paper we will study the extinguishing of the secondary wind by the accretion process.

\subsection{The delicate nature of the artificial viscosity}
\label{sec:viscosity}

\begin{figure}
 \centering

   \includegraphics[scale=0.95]{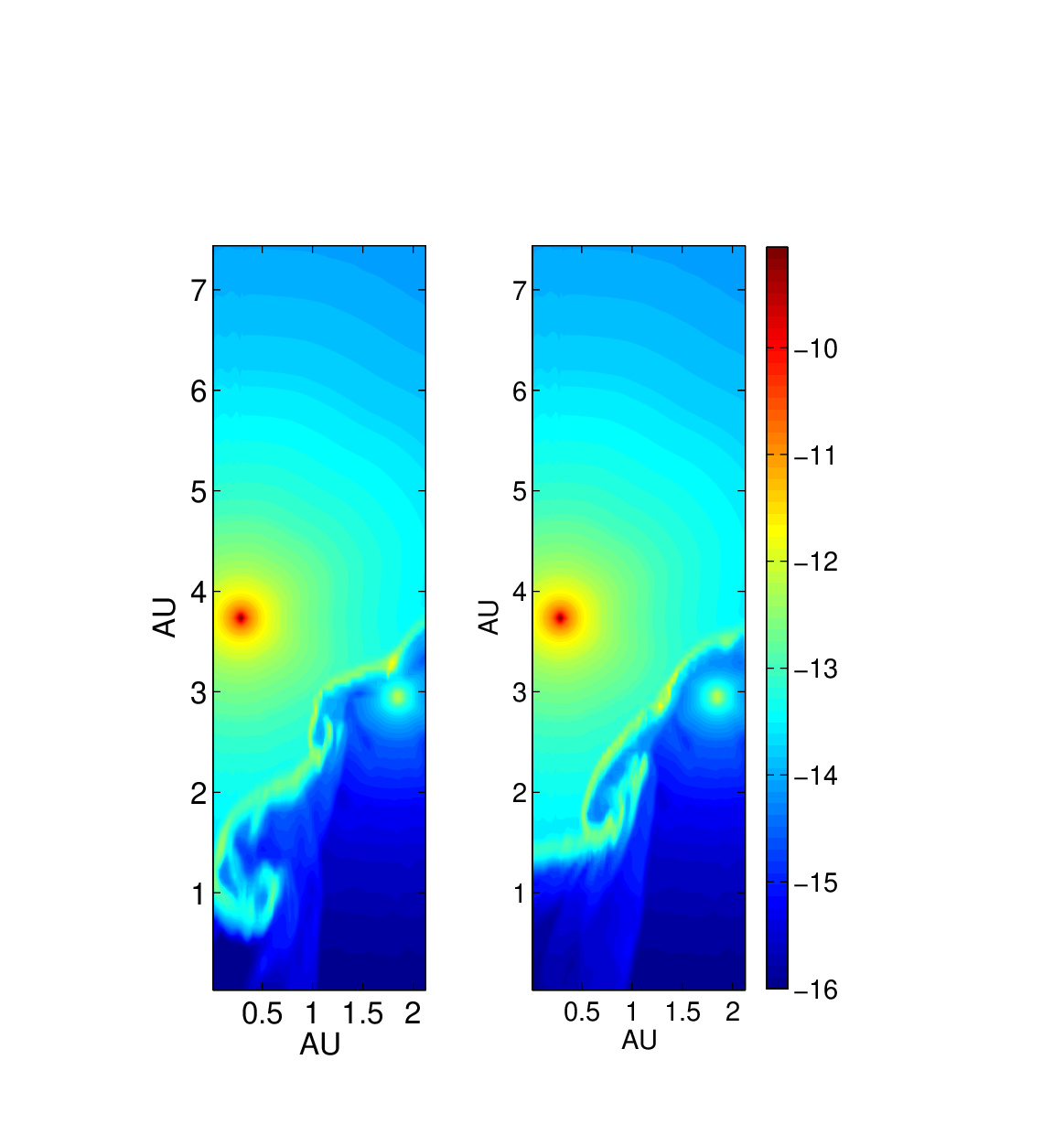}

        \caption{Comparing runs with regular and high artificial (numerical) viscosity in the $xy$ (equatorial) plane.
         Left: Density map of our standard run A2 that has a regular artificial viscosity.
         Right: Density map of run A4 that has an increased artificial viscosity, but other than that is identical
         to run A2. Both panels show the system -4.1 days relative to periastron passage.
         The bar on the right gives the log scale of density, and color-coded in units of $ \g \cm^{-3} $.}
   \label{fig:visc1}
     \end{figure}

\begin{figure}
 \centering

   \includegraphics[scale=0.95]{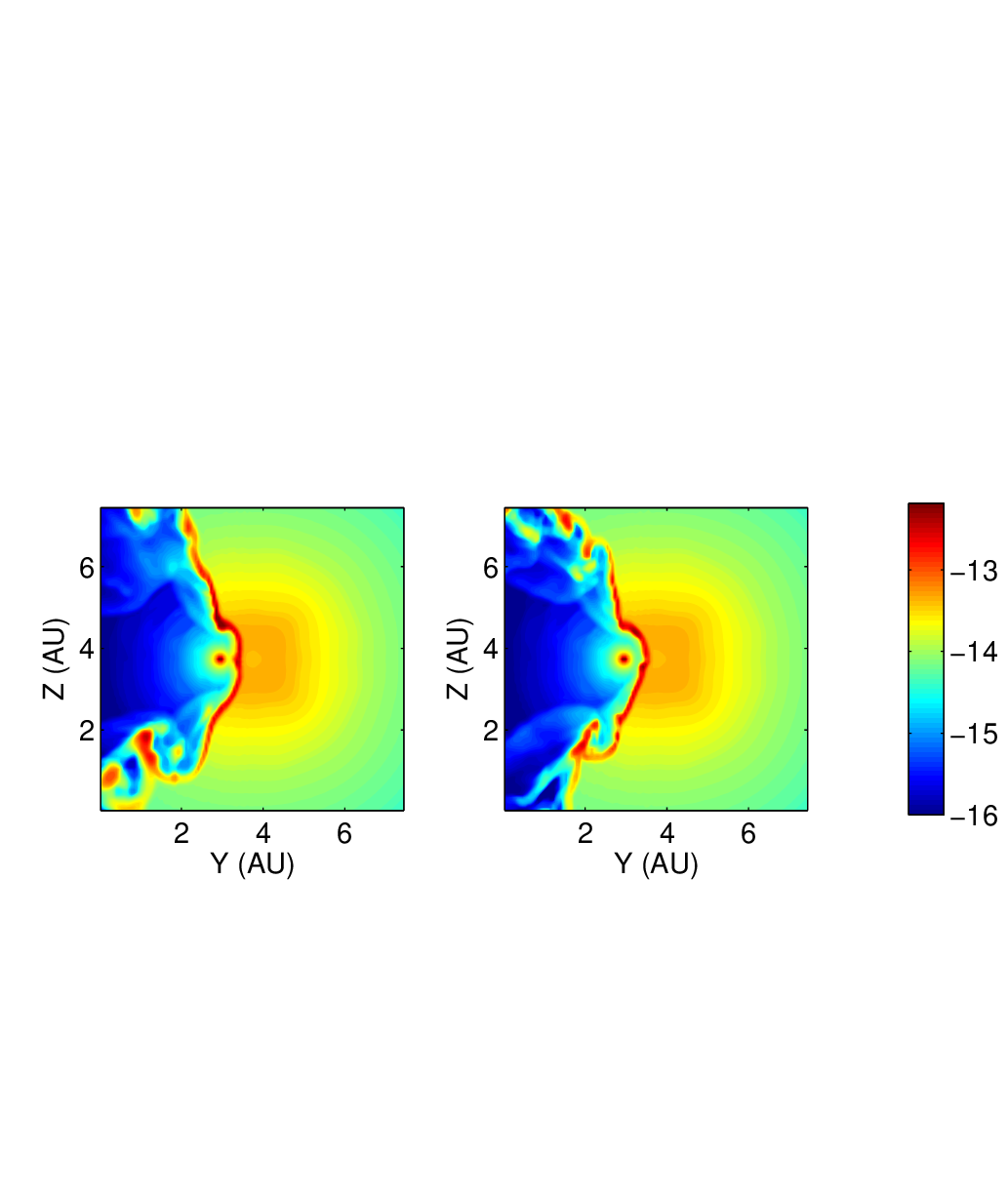}

        \caption{Like Fig. \ref{fig:visc1} but in the $yz$ plane.}
   \label{fig:visc2}
     \end{figure}

One should be aware of the artificial viscosity that is added to numerical codes to handle
discontinuities and sharp jumps of density and pressure.
The same artificial viscosity that prevents numerical instabilities in sharp discontinuities and shocks might
delay the growth of real physical instabilities.
To check that, we performed run A4 that has the same parameters as run A2, but
with increased artificial viscosity.
The comparison of the density profiles of the two runs is presented in Fig. \ref{fig:visc1} (the equatorial plane) and
in Fig. \ref{fig:visc2} (the $yz$ plane).

It is clear from the comparison that artificial viscosity can suppress the wiggling of the dense shell,
and of the formation of large dense clumps in the thin shell of the colliding winds.
As accretion is triggered by these large dense clumps,
we conclude that too high artificial viscosity might prevent accretion in such binary systems.
We cannot tell whether a too high artificial (numerical) viscosity prevented Parkin et al. (2011)
from finding accretion in their simulations. We here only raise the point that people conducting numerical simulations
of this kind should be aware of the role of artificial viscosity.

\subsection{The importance of periastron passage }
\label{sec:stand}

That accretion takes place in $\eta$ Car near periastron passage is not a trivial matter.
Several effects should operate together.
(1) The secondary accretion radius $2G M_2/v_1^2$ is not much smaller than the
distance from the secondary to the stagnation point of the winds collision region (Soker 2005).
This requires relatively slow primary wind, and massive secondary,
as well as close periastron passage.
(2) Dense primary wind that cools fast. This leads to the formation of a
dense thin shell, whose instabilities form dense clumps and lead to wiggling of the shell.
These are the dense clumps that trigger accretion.
(3) A large density ratio between the primary wind and the secondary wind, such that the dense clumps
are not pushed outward by the ram pressure (or radiation) of the secondary star (Soker 2005).

\begin{figure}
\centering

  \includegraphics[scale=0.85]{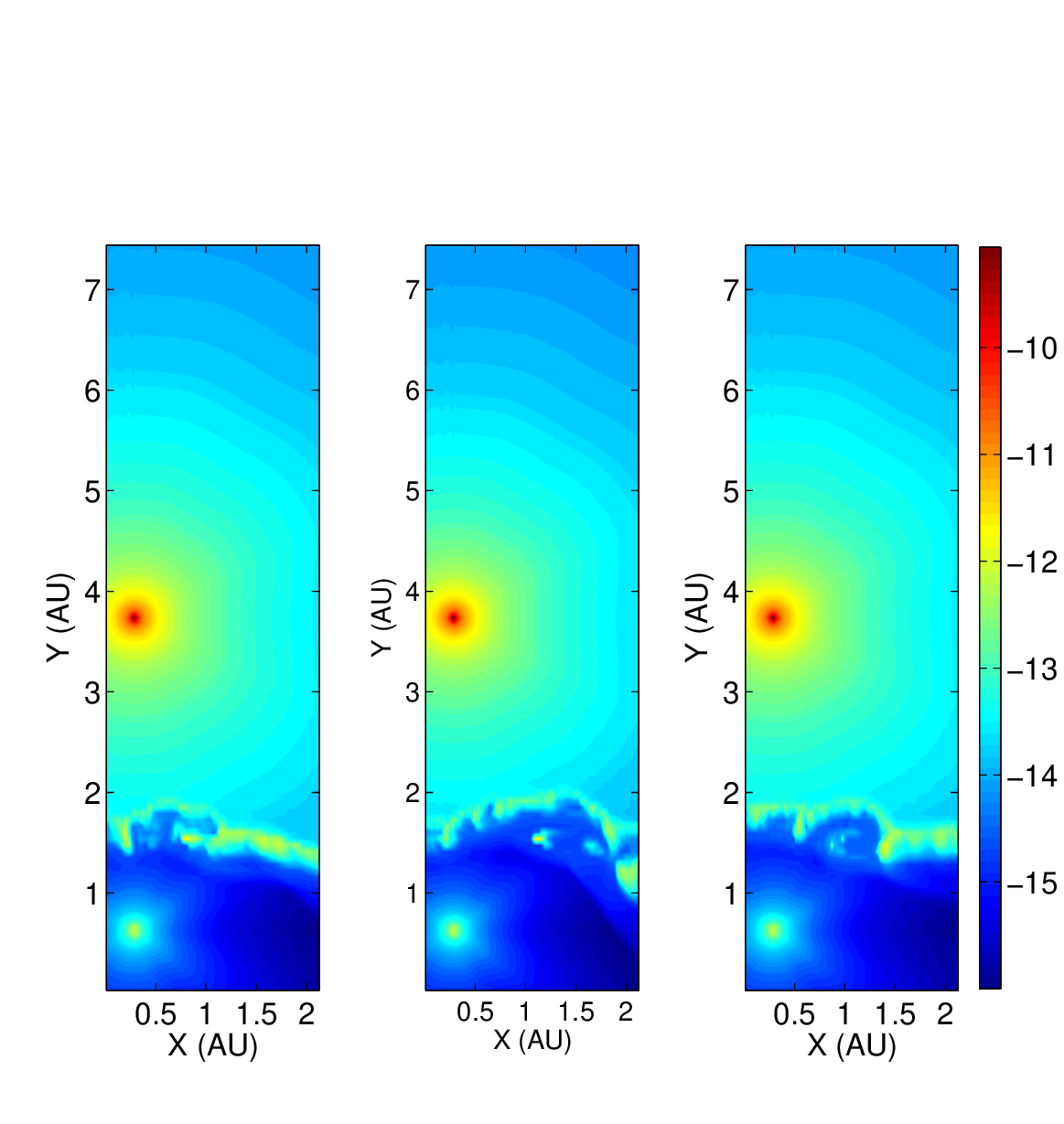}

       \caption{The density maps in the equatorial plane at three times for run A5, where the secondary is static;
       all other parameters as in run A2. These are to compare with Fig. \ref{fig:Standard1} of run A2.
       The left, middle and right panels are -15.6, -11 and -5.2 days relative to periastron passage, respectively.
       Accretion does not occur.
       It is evident that the motion of the secondary closer to the primary is required for accretion to begin.
       The bar on the right gives the log scale of the density, and color-coded in units of $ \g \cm^{-3} $.}
    \label{fig:stand}
    \end{figure}

We examine now the case where the secondary star does not come any closer to the
primary star, but rather stays static where we start the simulations, i.e., at a position angle of $90^\circ$ to
periastron, and at a distance of $3.17 \AU$ from the primary.
The purpose of this exercise is to demonstrate the importance of the orbital motion when accretion starts.
In Fig. \ref{fig:stand} we present the results of run A5, that has
parameters as in run A2, but the secondary does not move.
We show the flow at three times to be compared with run A2 in Fig. \ref{fig:Standard1}.
Even after continuing the run for a long time, no accretion takes place.
The thin shell suffers the instabilities: it wiggles inward and outward,
fluctuates, and dense clumps are formed.
Still, no accretion occurs as the clumps are at too large a distance from
the secondary star.

\subsection{The far side of the conical shell}
\label{sec:widegrid}

\begin{figure}
\centering

\includegraphics[scale=0.95]{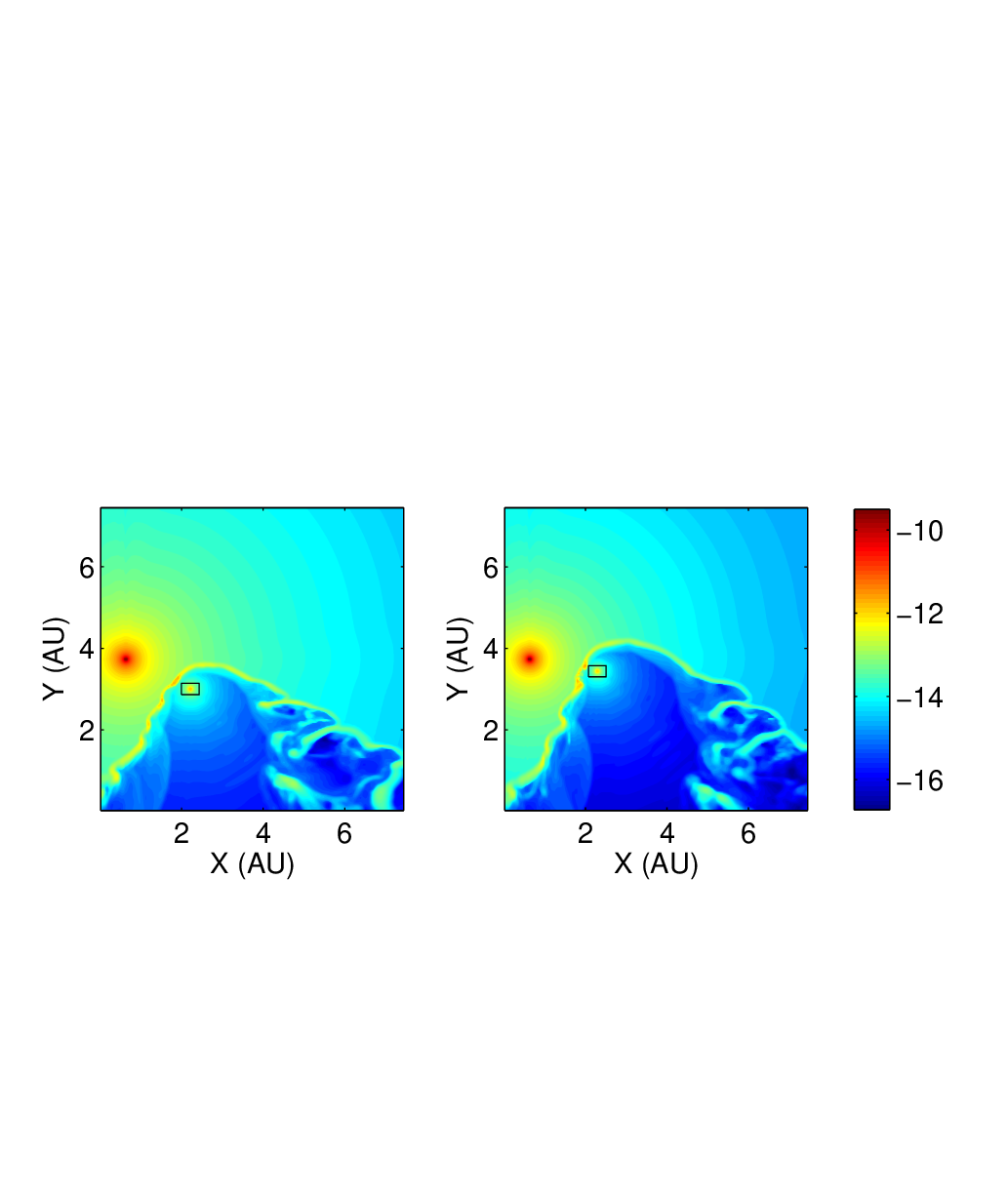}
       \caption{The flow structure on the far side of the winds interaction region in the equatorial plane
        ($xy$ plane) for the extended grid run A6. It has the same parameters as run A2,
        but the grid is enlarged in the X-direction.
        The left and right panels are -13.7 and -1.6 days relative to periastron passage, respectively.
        In the right panel accretion onto the secondary star takes place,
        in agreement with the results in the standard run A2 (Fig. \ref{fig:Standard1}).
        The zone inside rectangle is the injection region of the secondary star. It is wide in the x-axis since we still
        inject from a $11\times7\times7$ cells zone, and the cell size in the x direction is wider than before.
        The bar on the right gives the log scale of density, and color-coded in  units of $ \g \cm^{-3}$.}
   \label{fig:farside}
    \end{figure}

Parkin et al. (2011) results show that as the secondary approaches periastron
passage the far side of the conical shell displays much larger-amplitude perturbations
that the region between the two stars.
To examine our approach of starting the simulation at a binary angle
of $90^\circ$ from periastron, rather than from a much larger distance as in Parkin et al. (2011), we
extended the grid to follow the far side of the conical shell.
The results are displayed in Fig. \ref{fig:farside}.
We find that our simulations show very large amplitude instabilities in the
far side of the conical shell, and that its general shape is as obtained by Parkin et al. (2011).
The large-amplitude instabilities in the far side further ensure us that the viscosity in
the code is not too large (see section \ref{sec:viscosity}).

\section{SUMMARY}
\label{sec:summary}

We perform 3D hydrodynamical simulations of the winds interaction process in $\eta$ Carinae.
We include the gravity of the secondary star and the orbital motion starting 19 days
before periastron passage.
The results are presented in the different figures that show the density, temperature, and velocity maps
in different planes in the 3D computational grid.

Our main finding is that accretion of dense primary wind material by the secondary star
close to periastron passage in the $\eta$ Car binary system is inevitable.
Behind the onset of the mass accretion process are dense clumps (Akashi \& Soker 2010).
The interaction of the two winds lead to the formation of a thin dense shell
from the rapidly cooling shocked primary wind.
The thin shell suffers from instabilities that lead to its wiggling and the formation of dense clumps.
As the system approaches periastron the amplitudes of the instabilities increase, a phenomena
supported by observations of larger X-ray intensity fluctuations (Davidson 2002; Corcoran 2005; Moffat \& Corcoran 2009),
and the dense clumps come closer to the secondary star.
Accretion starts.
There is no long-lasting effects of the accreted mass.
This is because after several weeks the secondary wind rebuilds itself,
and within several months blows a mass that is larger than the accreted mass (Kashi \& Soker 2009a).

We perform several tests that teach us about the triggering of the accretion process.
{}From section \ref{sec:viscosity} we learn that to find an accretion in numerical simulations of systems
like $\eta$ Car, it is crucial not to have a too high numerical (artificial) viscosity.
The results, in particular of section \ref{sec:stand}, show that had the periastron distance been
larger by a factor of $\sim 2$ or more than its value in $\eta$ Car, accretion would not have occurred.
As well, if the primary wind was weaker, the dense shell would not be as dense, and
it would not approach close enough to the secondary to trigger accretion.
Over all, there are unique properties of the $\eta$ Car binary system that lead to accretion as the
binary stars approach periastron.
The accretion process was proposed (Soker 2005) to explain the deep minimum in the X-ray emission, and
was postulated to start several days before periastron passage.
Our results confirm this postulate: accretion starts about a week before periastron passage.

Some effects which are not included here are expected to further facilitate the accretion process:
\begin{enumerate}
\item Clumpy primary wind. The primary wind is thought to be clumpy (Moffat \& Corcoran 2009).
In our numerical simulations the primary wind is smooth, and the clumps that trigger the
accretion are formed in the thin dense shell.
The presence of dense clumps already in the preshock primary wind will make the accretion start somewhat
earlier and shut off the secondary wind more rapidly.
\item The acceleration zone of the primary wind. We initiate the winds at their terminal speeds.
Considering the acceleration zone of the primary wind, that might extend to a distance of
more than the periastron distance, would yield a higher accretion rate (Kashi \& Soker 2009a).
\item The acceleration zone of the secondary. As the clumps come closer to the secondary, the secondary radiation
acts to slow them down. We note two things.
First, this is not sufficient to prevent accretion (Soker 2005; Akashi \& Soker 2010).
Second, close to the secondary star the clumps enter the acceleration zone of the secondary wind.
In this region the ram pressure of the secondary wind is lower.
\item A higher mass loss rate into the primary wind will increase the accretion rate and its duration.
Groh et al. (2012) take a three times higher primary mass loss rate, and a somewhat lower velocity, $\dot M_1=8.5 \times 10^{-4} M_\odot \yr^{-1}$ and $v_1=420 \km \s^{-1}$,
(compared with our values of $\dot M_1=3 \times 10^{-4} M_\odot \yr^{-1}$ and $v_1=500 \km \s^{-1}$).
These values increase the primary wind density at each radius by a factor of $\sim 3$, hence facilitate accretion.
If, on the other hand, the primary wind density decreases (e.g., Mehner et al. 2010), then accretion might occur,
but it will last for a shorter time (Kashi \& Soker 2009b). 
\end{enumerate}

Over all, we estimate that the two contradicting effects, of radiation pressure and lower wind's ram pressure,
ensure that we are not in error in deducing the accretion process while neglecting the radiation pressure of he
secondary star.

Now that the accretion process explanation for the several weeks long X-ray deep decline is established,
the community should look for additional effects the accretion process has on other emission and absorption lines and
continuum bands across periastron passage.
In addition, higher resolution codes should resolve the accretion process and
study the shut-off process of the secondary wind by the accreted mass.
This is the subject of a forthcoming paper.

We thank an anonymous referee for helpful comments.
AK acknowledges a grant from the Irwin and Joan Jacobs fund at the Technion.
This research was supported by the Asher Fund for Space Research
at the Technion, and the Israel Science foundation.

\footnotesize

 \end{document}